# Dispersion Stability and Thermal Conductivity of Propylene Glycol Based Nanofluids

**Ibrahim Palabiyik, Zenfira Musina\*, Sanjeeva Witharana, Yulong Ding**

*Institute of Particle science and Engineering,*
*School of Process, Environmental and Materials Engineering, University of Leeds,*
*Woodhouse Lane, Leeds,UK, LS2 9JT*
*\*Corresponding author. e-mail: z.musina@leeds.ac.uk, tel: +44 (0)113 343 2543*

***Abstract*** The dispersion stability and thermal conductivity of propylene glycol based nanofluids containing $Al_2O_3$ and $TiO_2$ nanoparticles were studied in the temperature range of 20 to 80 °C. Nanofluids with different concentrations of nanoparticles were formulated by the two-step method without use of dispersants. In contrast to the common belief the average particle size of nanofluids was observed to decrease with increasing temperature. The nanofluids showed excellent stability over the temperature range of interest. Thermal conductivity enhancement for both of studied nanofluids was a non-linear function of concentration while was temperature independent. Theoretical analyses were performed using existing models and comparisons were made with experimental results. The model based on the aggregation theory appears to yield the best fit.

*Introduction*

Heat transfer processes play an important role in various areas of industrial engineering such as food and chemical processing, automobile, HVAC and power generation. Miniaturisation of devices and the demand for faster operation pronounces the need for quicker heat removal systems with enhanced heat transfer properties. Since proposed by Choi et al in 1995 (Choi et al 1995), nanofluids have attracted a considerable audience, as reflected by the exponentially increasing number of publications occurred in particular over the last 5 years. However, most of these studies have been carried out on aqueous-based nanofluids at the conditions close to the environmental temperatures. This formed the motivation for this work on the use of propylene glycol (PG) based formulations. PG has a much higher boiling point and a much lower freezing point in comparison with water. However PG has a lower thermal conductivity, about 1/3 of water. As a result there is a great potential to enhance the thermal conductivity of PG. This is one of the objectives of this work.

PG stands to displace ethylene glycol (EG) as a heat transfer fluid in applications where toxicity might be a concern, as in food processing etc. Despite its versatility, the use of PG as the base liquid has been rarely studied (Prasher et al. 2006b). This work aims to understand the stability and thermal conductivity of propylene glycol based $Al_2O_3$ and $TiO_2$ nanofluids. The thermal conductivity data obtained in present work are compared with different models and correlations available in the literature.

*Experimental*

Nanofluids were formulated by so-called two-step method as described elsewhere (Chen et al. 2007) by using propylene glycol (98%, Fluka), and $TiO_2$ (P25) and $Al_2O_3$ (size??Alu C) nanoparticles (both from Degussa, Germany). The densities of $TiO_2$ and $Al_2O_3$ nanoparticles were 4170 and 3970 kg/m$^3$ respectively. Nanofluid samples were prepared with nanoparticle concentrations of 1, 6 and 9 wt%, which is equal to $TiO_2$ 0.25, 1.56, 2.40 vol% and $Al_2O_3$, 0.26, 1.63, 2.51 vol% respectively.

Particle sizes of nanofluids were measured by using a Zetasizer Nano ZS (Malvern Instrument, UK) at 25 °C. Five consecutive data points were gathered for each measurement. The primary particle size of nanoparticles determined by transmission electron micrograph (TEM) (Philips CM200) was 21 nm for TiO2 and 13 nm for $Al_2O_3$.

In order to prepare stable nanofluids and reduce the size of agglomerates, long term sonication was applied using a Digital Sonicator (Model S70H by Elma, Germany). The average size of $TiO_2$ nanoparticles in PG was measured occasionally during sonication (Figure 1a). The initial average particle size of $TiO_2$ in PG was 520 nm, which after 16 hr of sonication was reduced to 115 nm. For $Al_2O_3$ nanofluids rapid particle size change up to constant values within 10 hr of sonication was observed. The nanofluids prepared as above without milling or using any surfactant/pH adjustment displayed excellent stability without any visible sedimentation for the several months.

Dispersion stability at different temperatures was evaluated as follows: The nanofluid samples were placed in an isothermal water bath and heated at 50 °C for 80 hr, followed by 70 °C and 80 °C consecutively for 40 hr each. At the end of each heating period, the particle size and pH were measured at 25 °C. Nanofluids with concentration of 6 and 9 wt% of nanoparticles were diluted to 1 wt% before the particle size measurement.

The thermal conductivity of the samples was measured with a Lambda Measuring System (PSL Systemtechnik GmbH, Germany) in a temperature range from 20 to 80 °C soon after 16 hr of sonication.. This instrument is based on transient heat-wire method that allows to measure thermal conductivity of nanofluids with minimum convectional effects.

*Results and Discussions*

*Stability of nanofluids*

Evaluation of the stability of the studied nanofluids at different temperatures was done by visual inspection and particle size monitoring. Titania and alumina nanofluids in PG showed excellent stability without any particles sedimentation for several months. However, according to TEM images, nanoparticles were not well dispersed and some agglomerates were present (Fig 1 b, c).

Earlier it was stated that linear or branched nanoclusters, such as shown here, provide the 'backbone' for faster heat conduction (Keblinski et al. 2002; Prasher et al. 2006b). It was speculated as a major reason for nanofluids to display thermal conductivity beyond Maxwell predictions. In order to characterize nanofluids stability particle size analysis was carried out. Although earlier it has been stated that higher temperatures increase the probability of particle aggregation and instability of nanofluids (Prasher et al. 2006a), our results for titania and alumina in PG clearly showed the decrease in particle sizes with increasing temperature (Figure 2). Thus, significant decreases (for more than 20 nm) were observed for 1 wt% and 6 wt% of PG-$TiO_2$ nanofluids. Same, for PG-$Al_2O_3$ samples the monotonous particle size change was discovered. Only for the sample containing 9 wt% of PG-$TiO_2$, particle size remains about 108 nm. Present observations prove that prolonged heating of the titania and alumina nanofluid did not cause particles aggregation.

*Thermal conductivity*

At the instrument calibration stage, measured thermal conductivity data obtained for pure PG were compared with literature data (Sun and Teja 2004). The difference was found to be less than 5% for all studied temperatures. Thermal conductivity enhancement was calculated from measured thermal conductivity data as follows:

$$Enhancement = (k-k_o)*100/k_o$$

where k and $k_o$ are the thermal conductivities of nanofluid and PG respectively.

As seen from Figure 3, the thermal conductivity enhancement increases non-linearly with nanoparticles concentration for both types of nanofluids. The present findings are consistent with other research (Yu et al. 2008; Yu et al. 2009) which supposed that at high volume concentration nanoparticles have a bigger effect on the viscosity of nanofluids than thermal conductivity enhancement. The largest overall thermal conductivity enhancement was 11% for $Al_2O_3$ and 9% for $TiO_2$. In comparison to titania, the alumina nanofluids have displayed superiority in performance, which might be explained by the higher thermal conductivity of alumina (40.0 W/mK, against 8.9 W/mK of titania) (Abu-Nada 2008).

A considerable number of models and correlations have been proposed in view of explaining the thermal conductivity behaviour of suspensions containing small solid particles. Some of those theories are analysed and compared with our experimental findings below. Maxwell was the first to propose the model to determine the effective electrical or thermal conductivity of suspensions containing solid particles (Maxwell 1881). This model may be applied to statistically homogeneous and low-volume fraction liquid–solid suspensions with randomly dispersed, uniformly sized, and non-interacting spherical particles:

$$\frac{k}{k_o} = \frac{k_p+2k_o-2\varphi(k_o-k_p)}{k_p+2k_0+\varphi(k_o-k_p)} \qquad (1)$$

where $k_p$, is the thermal conductivity of the nanoparticles and $\varphi$ is the volume fraction of nanoparticles in the mixture. Note that Maxwell equation is not a function of temperature.

Since then several models were introduced in order to take account of Brownian motion of nanoparticles (Koo and Kleinstreuer 2004; Patel et al. 2003), liquid layering around them (Wang et al. 2003), ballistic heat transport in nanoparticles and particle's geometry (Domingues et al. 2005; Evans et al. 2006; Lee 2007; Nie et al. 2008; Keblinski et al. 2002). However, it is largely accepted that the thermal conductivity enhancement can be best described by nanoparticle structuring (Ding et al. 2007) and the manner of particle packing in the matrix (Nielsen 1974). Thus, Hamilton and Crosser modified Maxwell's model and showed the effect of particle shape and particle volume fraction on thermal conductivity of suspensions (Hamilton and Crosser 1962). On the other hand, the size effects of nanoparticles are not included in both models. Considering that nanoparticles in nanofluids are mostly in the form of aggregates, Chen et al (Chen et al. 2009) used to modify the conventional form of Hamilton-Crosser model. It was introduced the concept of the effective volume fraction of aggregates $\varphi_a$ and replaced the term $k_p$ with $k_a$, which is the thermal conductivity of agglomerates as follows by:

$$\frac{k}{k_o} = \frac{k_a + 2k_o - 2\varphi_a(k_o - k_a)}{k_a + 2k_o + \varphi_a(k_o - k_a)}$$

(2)

Where $\varphi_a = (r_a/r)^{3-D}$ and $k_a$ is to be determined from the (Bruggeman 1935) model:

$$\frac{k_a}{k_o} = \frac{1}{4}\left\{(3\varphi_i - 1)\frac{k_p}{k_o} + (3(1-\varphi_i) - 1) + \left[\left((3\varphi_i - 1)\frac{k_p}{k_o} + (3(1-\varphi_i) - 1)\right)^2 + 8\frac{k_p}{k_o}\right]^{\frac{1}{2}}\right\}$$

(3)

Where $\varphi_i$ is the solid volume concentration of agglomerates given by $\varphi_i = (r_a/r)^{D-3}$, $r_a$ and r are respectively the radius of aggregates and primary nanoparticles. The term D is the fractal index, which has an average value of 1.8 for nanofluids assuming diffusion limited aggregation.

Moreover, some authors (Eapen et al. 2007; Keblinski et al. 2008; Shima et al. 2009) consider mean-field boundary theory as the best suited model for estimating thermal conductivity enhancement. There are two main models in mean field theory. One is the simple series and parallel model which is based on the configuration of nanoparticles relative to the direction of heat flux in a nanofluid (DeVera and Strieder 1977). According to the theory the effective thermal conductivity is calculated assuming series and parallel configuration of nanoparticles in base fluid is as follows:

$$k_{series} = \frac{(1-\varphi)}{k_o} + \frac{\varphi}{k_p}$$

(4)

$$\frac{1}{k_{parallel}} = (1-\varphi)k_o + \varphi k_p$$

(5)

Another widely used model in recent nanofluids literature is the Hashin–Shtrikman (HS) model which estimates upper and lower limits of effective thermal conductivity of nanofluids according to the formula below (Hashin and Shtrikman 1962):

$$k_o\left(1+\frac{3\varphi(k_p-k_o)}{3k_o+(1-\varphi)(k_p-k_o)}\right) \leq k \leq k_p\left(1-\frac{3(1-\varphi)(k_p-k_o)}{3k_p-\varphi(k_p-k_o)}\right)$$

(6)

Physically, the upper HS bound corresponds to a nanocluster matrix with spherical inclusions of fluid regions while the lower HS bound assumes well dispersed nanoparticles in the base fluid (Eapen et al. 2007). HS lower bound, which is the left hand side of the inequality, is identical to Maxwell's equation.

Thermal conductivity enhancement obtained from aforementioned models was compared with current experimental data for PG-$TiO_2$ and PG-$Al_2O_3$ nanofluids at 20 ˚C (Figures 4 and 5). The experimental data lies closer to the Chen's model than any other model. In an attempt to resolve the controversy of thermal conduction in nanofluids, Keblinski et al analysed a large number of experimental data published on nanofluids (Keblinski et al. 2008). It was demonstrated that almost all the literature data falls within HS bounds and showed that the well dispersed nanofluids follows the classical Maxwell relationship. Therefore, the thermal conductivities predicted by the Maxwell model are lower than the experimentally measured conductivities. As stated previously, studied there nanofluids were very stable. It can therefore be presumed that the aggregates were fairly well dispersed. If this assumption is true, the agreement of experimental data with Maxwell's model is further justified. HS model stands as a quite reasonable technique to estimate the thermal conductivity of nanofluids. HS model is more accurate and defines narrower range than simple series and parallel model for thermal conductivity enhancement of nanofluids. In fact it is hard to control the nanoparticle structure and configuration with the existing knowhow and technology. This restricts the series-parallel model from becoming a viable technique. Present results confirm previous observations.

Maxwell model underestimates thermal conductivity enhancement for both $TiO_2$ and $Al_2O_3$ nanoparticles. Besides, Chen's model estimation for thermal conductivity enhancement ratio is very different for same volume concentration of two types of nanoparticles. On the other hand, Maxwell estimation is more or less the same for the above case. For instance, at 9 wt% of alumina and titania nanofluids, Maxwell model estimates thermal conductivity enhancement as ~5% for both fluids. However, for the Chen's model thermal conductivity enhancement is 35% for $TiO_2$ and 10% for $Al_2O_3$ samples. Moreover, the series and parallel models appear to be the most ineffective estimation tools for both nanofluids.

### *Effect of temperature on thermal conductivity*

Figure 6 presents the thermal conductivity enhancement as a function of measuring temperature. The thermal conductivity of the base liquid and nanofluids remains nearly constant in all studied temperature interval. It should also be noted that, the data of PG is normalised to its thermal conductivity value at 20 °C. Hence, it can be deduced that the thermal conductivity increase of nanofluids is not temperature dependent but follows the behaviour of the base fluid. Therefore, the thermal conductivity increases of nanofluids with temperature in the literature are essentially resulting from the base fluid properties. These findings are in good agreement with recent literature (Buongiorno at al JAP (106) 2009), Keblinski et al. 2008; Timofeeva et al. 2007).

*Conclusions*

This study was focussed on the experimental and theoretical analyses of propylene glycol based titania and alumina nanofluids. Stable nanofluids of 1, 6 and 9 wt% were prepared by prolonged ultrasonication without use of surfactants. Dispersion stability of the samples was examined by visual inspection and particle size measurements. In contrast to common belief, the average particle size of nanofluids decreased with gradually increasing temperature from 20 to 80 °C. $Al_2O_3$ and $TiO_2$ based nanofluids showed the same thermal conductivity behaviour as a pure PG. Thermal conductivity of the samples displayed a non-linear increase with the particle concentration. The enhancement of thermal conductivity was not temperature dependant. All experimental data for thermal conductivity were in good agreement with Chen's aggregation model (Chen et al. 2009). The present observations underline the capability of aggregation mechanism to accurately predict the thermal conductivity of well-dispersed nanofluids even at fairly high particle concentrations.

*Acknowledgements*

The authors thank UK EPSRC for supporting this research under EP/F000464/1 and P/F023014/1.

List of Figures

Figure 1a. Average particle size as a function of ultrasonication of 1 wt % PG-TiO2 nanofluids.
Figure 1b. TEM images of PG-$Al_2O_3$ nanofluids.
Figure 1c. TEM images of PG-$TiO_2$ nanofluids.
Figure 2. Change in average particle size with temperature in PG-$TiO_2$ and PG-$Al_2O_3$ nanofluids.
Figure 3. Thermal conductivity versus concentration for $TiO_2$ and $Al_2O_3$ nanofluids in PG at 20 °C.
Figure 4. Thermal conductivity enhancement versus concentration for PG-$TiO_2$ samples at 20 °C.
Figure 5. Thermal conductivity enhancement versus volume concentration for PG-$Al_2O_3$ samples at 20 °C.
Figure 6. Thermal conductivity enhancement versus temperature for $TiO_2$ and $Al_2O_3$ nanofluids. The data of PG is normalised to its thermal conductivity value at 20 °C whereas the data of nanofluids are normalised to thermal conductivity value of PG at the same temperature.

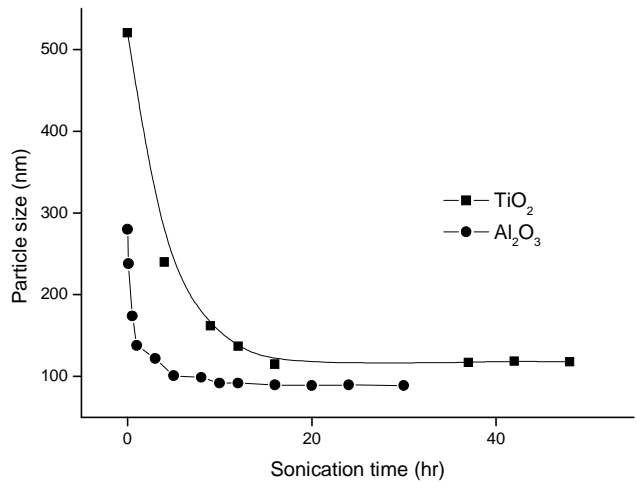

Figure 1a.

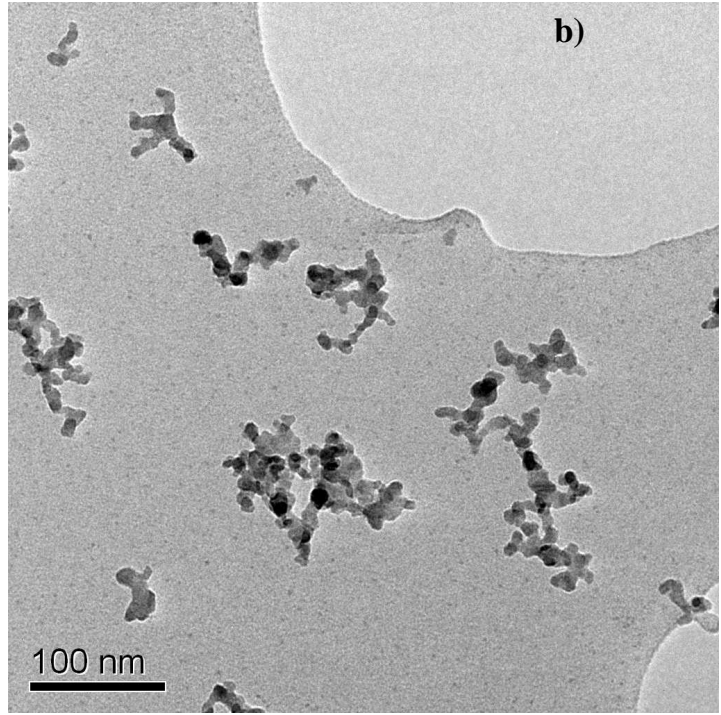

Figure 1b.

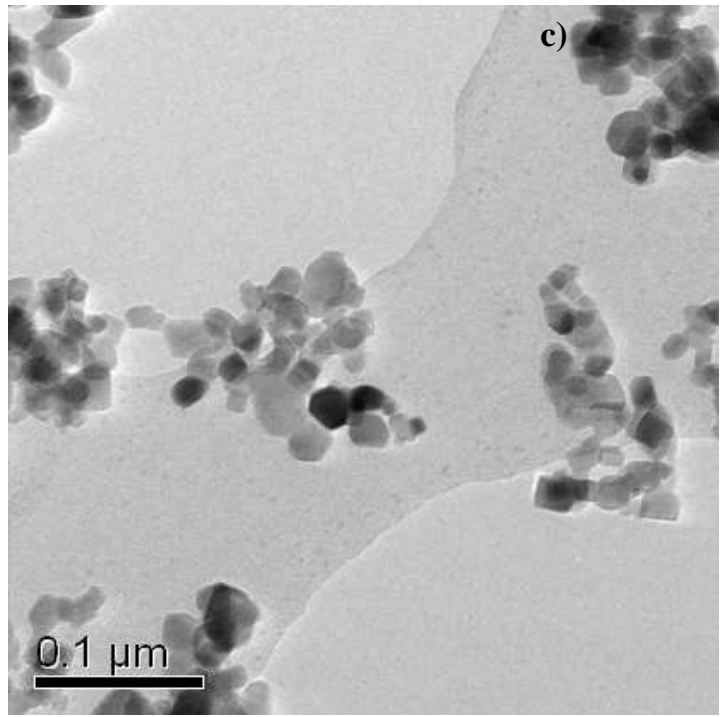

Figure 1c.

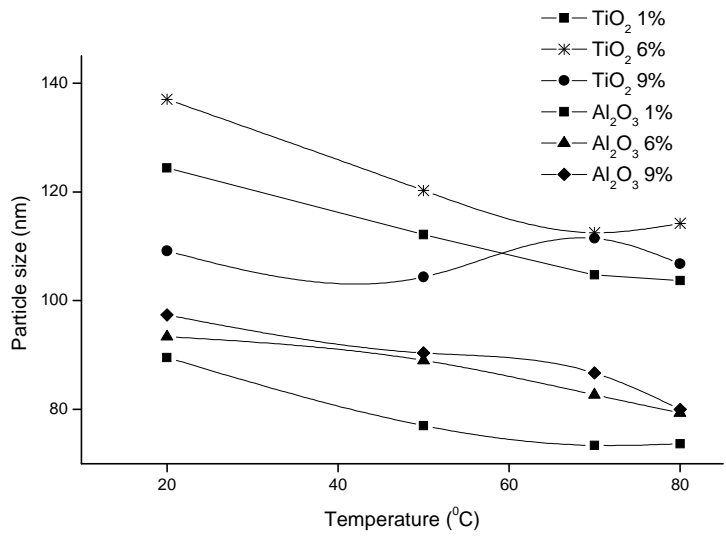

Figure 2.

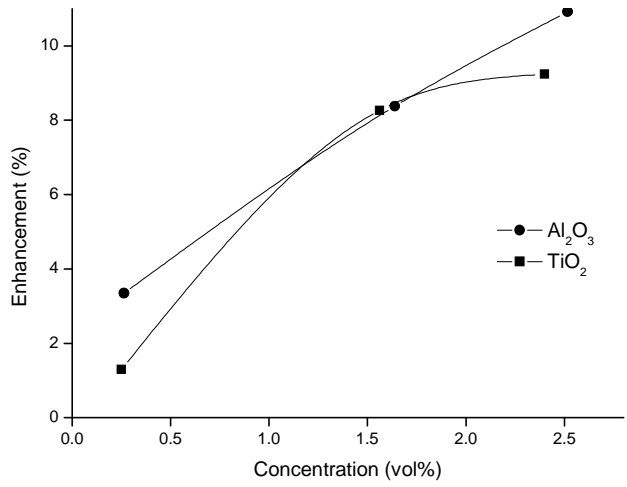

Figure 3.

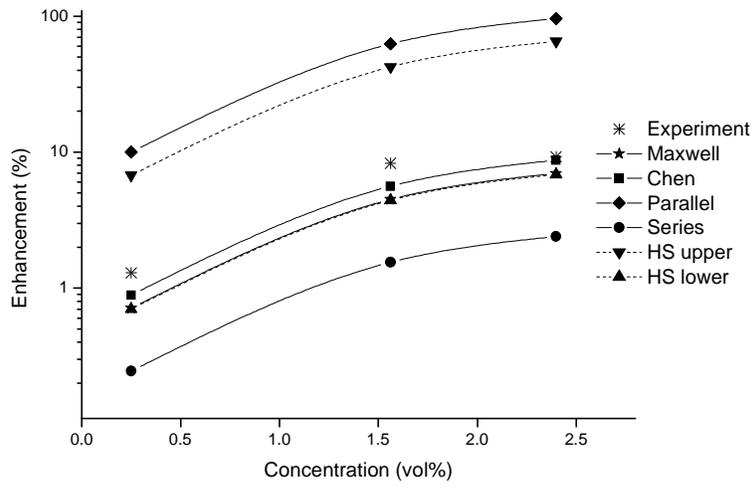

Figure 4.

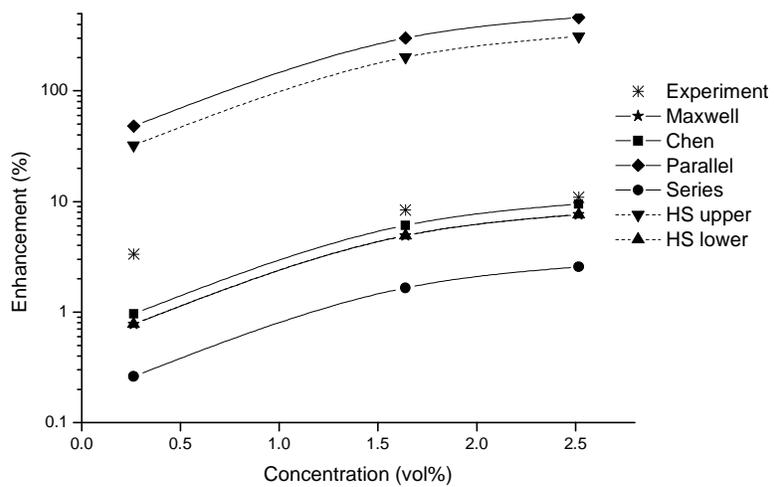

Figure 5.

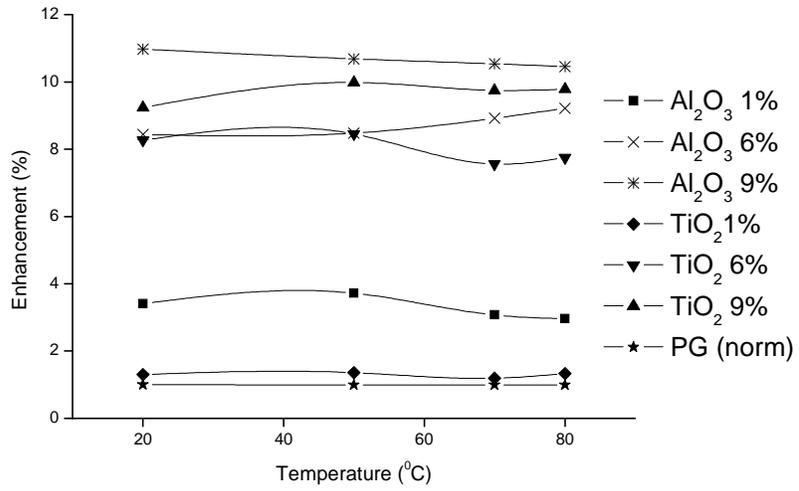

Figure 6.